\documentclass{article}
\usepackage{amsmath}
\usepackage{cite}
\usepackage{geometry}

\input{tcilatex}
\begin{document}

\title{Fourier, Gauss, Fraunhofer, Porod and the Shape from Moments Problem}
\author{Gregg M. Gallatin \\
National Institute of Standards and Technology\\
Center for Nanoscale Science and Technology\\
Gaithersburg, MD\\
gregg.gallatin@nist.gov}
\maketitle

\begin{abstract}
We show how the Fourier transform of a shape in any number of dimensions can
be simplified using Gauss's law and evaluated explicitly for polygons in two
dimensions, polyhedra three dimensions, etc. We also show how this
combination of Fourier and Gauss can be related to numerous classical
problems in physics and mathematics. Examples include Fraunhofer diffraction
patterns, Porods law, Hopfs Umlaufsatz, the isoperimetric inequality and
Didos problem. We also use this approach to provide an alternative
derivation of Davis's extension of the Motzkin-Schoenberg formula to
polygons in the complex plane.
\end{abstract}

\section{Introduction}

A shape can be defined mathematically in many ways. Here we consider
defining it by a function that has the value unity inside the shape and zero
outside the shape. Only simply connected orientable\cite{SCO} shapes will be
considered. The Fourier tranform of a shape defined in this way has
interesting applications and connections to many areas of physics and
mathematics.\ For example, in physical optics the Fourier transform of the
shape of an aperture or opening in an opaque screen yields the Fraunhofer or
far field diffraction pattern that is generated when light passes through
the aperture\cite{FourierOptics}. In X-ray scattering, the Fourier transform
of a volumetric shape in three dimensions reduces in the appropriate limit
to Porods law\cite{Porod}. From a probabilistic point of view the Fourier
transform of a shape properly normalized can be considered as the
characteristic function or moment generating function of the shape. Hence
the Fourier transform of the shape is intimately related to the moments of
the shape and hence to the "shape from moments problem", which has been
studied recently for polygons by Golub, Milanfar and others\cite{Milanfar}%
\cite{SIAM} and is related to the overall problem of pattern recognition.
The Fourier transform of an area bounded by a smooth curve or a polygon in
the plane can be shown to imply Hopf's Umlaufsatz\cite{Hopf}, which states
that the tangent vector (or the normal vector or any linear combination
thereof) to a smooth or piecewise smooth closed orientable simply connected
curve in the plane rotates by $\pm 2\pi $ as it goes completely around the
curve. The $+/-$ signs correspond to counterclockwise/clockwise rotation.
This seemigly obvious geometrical fact is rather subtle to prove\cite{Hopf}
\ We also use this formula to derive the planar version Stokes law\cite%
{Stokes} and to show that a circle maximizes the area for a given perimeter
length which is known as the isoperimetric inequality\cite{Hopf}\cite%
{Isoperimetric}. Finally there is the famous problem faced by Dido, Queen of
Carthage, which was to determine what shape open curve of a given length
encloses the maximum area when its endpoints are connected by a straight
line with an arbitrary length\cite{Dido}. Here I\ show how\ combining the
Fourier transform with Gauss's Law provides a central platform for relating
and solving all these problems.

As an aside, the Fourier transform is just one of many techniques that are
used for shape discrimination\cite{SD}. This problem is both mathematically
interesting, and, given the ubiquity of digital data bases, technologically
important. Here our specific interest is in the relation of the Fourier
transform of a shape to its moments and to the problems mentioned above.
This is related to, but generally different from, the issues concerned with
using Fourier transforms explicitly for shape discrimination in digital data
bases and we will not discuss this aspect further.

The paper is organized as follows. In the Section 2, Gauss's Law is used to
rewrite the Fourier transform of a $D$ dimensional volume as the surface
integral over the $D-1$ surface of that volume. This suface integral can
then be evaluated exactly for polygons in two dimensions, polyhedra in three
dimensions, etc. In particular in two dimensions this then provides an
explicit relation between the vertices of a polygon and its moments. This
section also discusses how this result is related to the isoperimetric
inequality, Didos problem and Hopf's Umlaufsatz. Finally we show how the
combination of Fourier and Gauss provides an alternative derivation of a
famous result of Davis concerning the integral of an analytic fuction over a
polygon in the complex plane\cite{Davis}\cite{Milanfar}. Section 3 briefly
discusses how the solution for two dimensional polyhedra appears when
Fourier transforming three dimensional polyhedrons before showing how the
surface integral form of the three dimensional Fourier transform can be used
to derive not only the standard Porods law\cite{Porod} for X-ray scattering
from spherical particles but also the extension of this law to anisotropic
particles. This extension has been discussed recently in a series of papers
by Ciccariello and colleagues\cite{Ciccariello}. The derivation presented
here is somewhat more direct than that used by Ciccariello and colleagues.

\section{Fourier and Gauss}

For generality we begin in $D$ dimensions but will rapidly particularize to
2 and 3 dimensions. Define a shape, in $D$ dimensions, by the function $%
\theta _{V}\left( \vec{x}\right) $ where $\theta _{V}\left( \vec{x}\right)
=1 $ for $\vec{x}=\left( x_{1},x_{2},\cdots ,x_{D}\right) $ inside $V$ and 0
for $\vec{x}$ outside. This is known in certain circles as an indicator
function\cite{IndicatorFunction}. Here we take $x_{i}$ for $i=1,2,\cdots ,D$
to be Cartesian coordinates. The (normalized)\ moments $\,\left\langle
x_{1}^{p_{1}}\cdots x_{D}^{p_{m}}\right\rangle $ of the shape are then given
by 
\begin{equation}
\,\left\langle x_{1}^{p_{1}}\cdots x_{D}^{p_{m}}\right\rangle =\frac{1}{v}%
\int d^{D}x~\theta _{V}\left( \vec{x}\right) ~x_{1}^{p_{1}}\cdots
x_{D}^{p_{D}}\equiv \frac{1}{v}\int\limits_{V}d^{D}x\;x_{1}^{p_{1}}\cdots
x_{D}^{p_{D}}  \label{1}
\end{equation}%
where $\int_{V}=\int \theta _{V}$ indicates integration over the shape, $%
v=\int d^{D}x~\theta _{V}\left( \vec{x}\right) $ is the finite volume of the
shape and $p_{i}=0,1,2,3,...$.for each $i.$ The symbols $V$ and $v$ are used
to distinguish between the shape itself and its volume as a numerical
value.\ Also the terms "volume" and "surface" are used generically and
should be understood to refer to a $D$ dimensional submanifold in $D$
dimensions and to its $D-1$ dimensional boundary, respectively.

Note that $\frac{1}{v}\theta _{V}\left( \vec{x}\right) $ can be thought of
as a probability density for $\vec{x}$ to be inside the shape. The
characteristic function or moment generating function $\bar{\phi}\left( \vec{%
\beta}\right) $ is then given by the Fourier transform $\theta _{V}\left( 
\vec{x}\right) /v$, i.e.,%
\begin{equation}
\bar{\phi}\left( \vec{\beta}\right) =\frac{1}{v}\int\limits_{V}d^{D}x~\theta
_{V}\left( \vec{x}\right) \;e^{i\vec{\beta}\cdot \vec{x}}  \label{2}
\end{equation}%
with $\vec{\beta}\cdot \vec{x}=\sum_{i=1}^{D}\beta _{i}x_{i}\equiv \beta
_{i}x_{i}.$ (Unless noted otherwise the Einstein summation convention,
wherein repeated indices are summed over their appropriate range, will be
used throughout the paper.) The bar on $\phi $ is meant to indicate that
this is the normalized charactistic function, i.e., $\bar{\phi}\left(
0\right) =1.$ Below we will find it convenient to work with the unnormalized
function $\phi \left( \vec{\beta}\right) $.

Using the power series representation of $\exp \left[ i\vec{\beta}\cdot \vec{%
x}\right] $ and the multinomial theorem it follows that%
\begin{align}
\bar{\phi}\left( \vec{\beta}\right) & =\frac{1}{v}\int\limits_{V}d^{D}x%
\sum_{n=0}^{\infty }\frac{\left( i\beta _{i}x_{i}\right) ^{n}}{n!}  \notag \\
& =\sum_{n=0}^{\infty }\frac{i^{n}}{n!}\sum_{p_{1}+p_{2}+\cdots +p_{D}=n}%
\frac{n!}{p_{1}!p_{2}!\cdots p_{D}!}\beta _{1}^{p_{1}}\cdots \beta
_{D}^{p_{D}}\left\langle x_{1}^{p_{1}}\cdots x_{D}^{p_{D}}\right\rangle
\label{3}
\end{align}%
This result indicates the simple but direct relation that exists between the
Fourier transform of a shape and its moments $\left\langle
x_{1}^{p_{1}}\cdots x_{D}^{p_{D}}\right\rangle $. Note that we have assumed
that since $v$ is finite we can interchange orders of integration and
summation. For all practical cases this is certainly true.

We see from this result that if we can compute $\bar{\phi}\left( \vec{\beta}%
\right) $ explicitly then the moments $\left\langle x_{1}^{p_{1}}\cdots
x_{D}^{p_{D}}\right\rangle $ follow directly from a Taylor expansion of $%
\bar{\phi}\left( \vec{\beta}\right) $ about $\vec{\beta}=0.$ An explicit
form for $\bar{\phi}\left( \vec{\beta}\right) $ can be found by applying
Gauss law to the Fourier integral. Noting that%
\begin{equation}
e^{i\vec{\beta}\cdot \vec{x}}=\vec{\partial}\cdot \left( \frac{\vec{\beta}}{%
i\beta ^{2}}e^{i\vec{\beta}\cdot \vec{x}}\right)  \label{4}
\end{equation}%
with $\vec{\partial}=\left( \partial /\partial x_{1},\partial /\partial
x_{2},\cdots ,\partial /\partial x_{D}\right) \equiv \left( \partial
_{1},\partial _{2},\cdots ,\partial _{D}\right) $ we get 
\begin{align}
\int\limits_{V}d^{D}xe^{i\vec{\beta}\cdot \vec{x}}& =\int\limits_{V}d^{D}x\;%
\vec{\partial}\cdot \left( \frac{\vec{\beta}}{i\beta ^{2}}e^{i\vec{\beta}%
\cdot \vec{x}}\right)  \notag \\
& =\frac{\vec{\beta}}{i\beta ^{2}}\cdot \int\limits_{\partial V}d^{D-1}s\;%
\sqrt{g\left( \vec{s}\right) }\hat{n}\left( \vec{s}\right) e^{i\vec{\beta}%
\cdot \vec{R}\left( \vec{s}\right) }  \label{5}
\end{align}%
Here $\partial V$ indicates the surface of $V$ with $\vec{s}=\left(
s_{1},s_{2},\cdots ,s_{D-1}\right) $ being coordinates on the surface $%
\partial V$. $\vec{R}\left( \vec{s}\right) $ gives the position in $D$
dimensions of the point on the surface labeled by $\vec{s}$ and 
\begin{equation}
g\left( \vec{s}\right) =\left\vert \det \left[ g_{ij}\left( \vec{s}\right) %
\right] \right\vert =\left\vert \det \left[ \partial _{s_{i}}\vec{R}\left( 
\vec{s}\right) \cdot \partial _{s_{j}}\vec{R}\left( \vec{s}\right) \right]
\right\vert  \label{6}
\end{equation}%
where $g_{ij}\left( \vec{s}\right) $ is the induced metric on $\partial V$, $%
\hat{n}\left( \vec{s}\right) $ is the local outward normal to the surface $%
\partial V$ and $\left\vert \cdots \right\vert $ indicates the absolute value%
\cite{Pearson}$.$Below we use (5)\ to prove that $\sqrt{\left\vert \det %
\left[ \partial _{s_{i}}\vec{R}\left( \vec{s}\right) \cdot \partial _{s_{j}}%
\vec{R}\left( \vec{s}\right) \right] \right\vert }d^{2}s$ is the area
element of a parallelogram in $2$ dimensions with sides defined by the
vectors $d\vec{R}_{i}=\partial _{s_{i}}\vec{R}\left( \vec{s}\right) ds_{i}$
(no sum on $i)$.and that\ $\sqrt{\left\vert \det \left[ \vec{a}_{i}\cdot 
\vec{a}_{j}\right] \right\vert }$ is the volume of the parallelogram formed
by 3 noncoplanar (nonzero) vectors $\vec{a}_{i},$ for $i=1,2,3$\ in three
dimensions and so $\sqrt{\left\vert \det \left[ \partial _{s_{i}}\vec{R}%
\left( \vec{s}\right) \cdot \partial _{s_{j}}\vec{R}\left( \vec{s}\right) %
\right] \right\vert }d^{3}s$ with $i,j=1,2,3$ is the volume element in three
dimensions.

Combining (3) and (5) yields 
\begin{eqnarray}
&&\frac{1}{v}\frac{\vec{\beta}}{i\beta ^{2}}\cdot \int\limits_{\partial
V}d^{D-1}s\;\sqrt{g\left( \vec{s}\right) }\hat{n}\left( \vec{s}\right) \cdot
e^{i\vec{\beta}\cdot \vec{R}\left( s\right) }  \notag \\
&=&\sum_{n=0}^{\infty }\frac{i^{n}}{n!}\sum_{p_{1}+p_{2}+\cdots +p_{m}=n}%
\frac{n!}{p_{1}!p_{2}!\cdots p_{D}!}\beta _{1}^{p_{1}}\cdots \beta
_{D}^{p_{D}}\left\langle x_{1}^{p_{1}}\cdots x_{D}^{p_{D}}\right\rangle
\label{7}
\end{eqnarray}%
It will be convenient below to work with the unnormalized form of the
characteristic function 
\begin{equation}
\phi \left( \vec{\beta}\right) =v\bar{\phi}\left( \vec{\beta}\right)
\label{8}
\end{equation}%
and write the unnormalized moments as%
\begin{equation}
M\left( x_{1}^{p_{1}}\cdots x_{D}^{p_{D}}\right) =v\left\langle
x_{1}^{p_{1}}\cdots x_{D}^{p_{D}}\right\rangle =\int d^{D}x~\theta
_{V}\left( \vec{x}\right) ~x_{1}^{p_{1}}\cdots x_{D}^{p_{D}}  \label{9}
\end{equation}%
Before particularizing to 2 and 3 dimensions note that expanding the left
hand side of $\left( 7\right) $ in powers of $\beta =\left\vert \vec{\beta}%
\right\vert $ yields a term proportional to $1/i\beta .$ This term must
vanish since there is no corresponding power of $\beta $ on the right and we
get the result that 
\begin{equation}
\int\limits_{\partial V}d^{D-1}s\;\sqrt{g\left( \vec{s}\right) }\hat{n}%
\left( \vec{s}\right) =0  \label{10}
\end{equation}%
This can be interpreted as the statement that "the boundary of a boundary is
zero". This boundary of a boundary principle has been considered to have
interesting implications with respect to the origin of physical laws.\cite%
{Boundary}

The remainder of the paper considers particular cases in 2 or 3 dimensions
where $\phi \left( \vec{\beta}\right) $ can be computed from the surface
integral either exactly or approximately. We now relate these results to
various "classical" problems in physics and mathematics.

\section{Results in Two Dimensions}

In this section we show how (5) and (7) can be used to derive various two
dimensional classical results\ in physics and mathematics.

\subsection{Smooth Curves}

For smooth curves in two dimensions (5) reduces to 
\begin{equation}
\int\limits_{V}d^{2}xe^{i\vec{\beta}\cdot \vec{x}}=\frac{\vec{\beta}}{i\beta
^{2}}\cdot \int\limits_{\partial V}ds~\hat{n}\left( s\right) e^{i\vec{\beta}%
\cdot \vec{R}\left( s\right) }  \label{11}
\end{equation}%
where $V$ is the region enclosed by the curve $\partial V$ given by $\vec{R}%
\left( s\right) $ with the parameter $s$ being the length along the curve.
With $ds$ defined as the unit of length on the curve, i.e., $ds^{2}=d\vec{R}%
^{2}$, it follows that $g\left( s\right) =1.$ The unit tangent vector to the
curve at $s$ is given by 
\begin{equation}
\hat{t}\left( s\right) =\partial _{s}\vec{R}\left( s\right)  \label{12}
\end{equation}%
which is automatically normalized since $ds=\sqrt{d\vec{R}^{2}},$ i.e.,$%
\left\vert \partial _{s}\vec{R}\left( s\right) \right\vert =1.$ (Generally
we will use a hat "\symbol{94}" to indicate the vector has been normalized
to have unit length. Also we will switch back and for between vector and
component notation, e.g., between writing $\hat{u}_{1}$ and $\left(
1,0\right) $.)

We take increasing $s$ to correspond to counterclockwise circulation of the
curve so that 
\begin{equation}
\hat{n}\left( s\right) =\varepsilon \cdot \hat{t}\left( s\right)
=\varepsilon \cdot \partial _{s}\vec{R}\left( s\right)  \label{13}
\end{equation}%
where 
\begin{equation}
\varepsilon =%
\begin{bmatrix}
0 & 1 \\ 
-1 & 0%
\end{bmatrix}
\label{14}
\end{equation}%
and the "$\cdot "$ indicates matrix multiplication, i.e.,%
\begin{equation}
n_{i}\left( s\right) =\varepsilon _{ij}n_{j}\left( s\right)  \label{15}
\end{equation}%
with repeated indices summed over 1,2. Note that $\varepsilon $ is the $D=2$
version of the totally antisymmetric tensor, often called the Levi-Civita
tensor\cite{Baez}, defined as +1 for even permutations of $i,j=1,2$, $-1$
for odd permutations and 0 if $i=j.$

Substituting (13 )\ into (11) and integrating by parts yields%
\begin{eqnarray}
\int\limits_{V}d^{2}xe^{i\vec{\beta}\cdot \vec{x}} &=&-\frac{1}{i\beta ^{2}}%
\int\limits_{\partial V}ds\;\left( \vec{\beta}\cdot \varepsilon \cdot \vec{R}%
\left( s\right) \right) \left( i\vec{\beta}\cdot \partial _{s}\vec{R}\left(
s\right) \right) e^{i\vec{\beta}\cdot \vec{R}\left( s\right) }  \notag \\
&=&-\int\limits_{\partial V}ds\;\left( \hat{\beta}\cdot \varepsilon \cdot 
\vec{R}\left( s\right) \right) \left( \hat{\beta}\cdot \partial _{s}\vec{R}%
\left( s\right) \right) e^{i\vec{\beta}\cdot \vec{R}\left( s\right) }
\label{16}
\end{eqnarray}%
The nominal $1/\beta $ pole in (11) has been cancelled and we can set $\beta 
$ to zero for arbitrary nonzero $\hat{\beta}=\vec{\beta}/\left\vert \vec{%
\beta}\right\vert $ which gives 
\begin{equation}
v=\int\limits_{V}d^{2}x=-\int\limits_{\partial V}ds\;\left( \hat{\beta}\cdot
\varepsilon \cdot \vec{R}\left( s\right) \right) \left( \hat{\beta}\cdot
\partial _{s}\vec{R}\left( s\right) \right)  \label{17}
\end{equation}%
Letting $\hat{\beta}$ be $\left( 1,0\right) $ or $\left( 0,1\right) $ with $%
\vec{R}\left( s\right) =\left( R_{1}\left( s\right) ,R_{2}\left( s\right)
\right) $ yields 
\begin{equation}
v=-\int\limits_{\partial V}ds\;R_{2}\left( s\right) \partial _{s}R_{1}\left(
s\right) =\int\limits_{\partial V}ds\;R_{1}\left( s\right) \partial
_{s}R_{2}\left( s\right)  \label{18}
\end{equation}%
Averaging the two forms gives%
\begin{eqnarray}
v &=&\frac{1}{2}\int\limits_{\partial V}ds\;\left( R_{1}\left( s\right)
\partial _{s}R_{2}\left( s\right) -R_{2}\left( s\right) \partial
_{s}R_{1}\left( s\right) \right)  \notag \\
&=&\frac{1}{2}\int\limits_{\partial V}ds\;R_{i}\left( s\right) \varepsilon
_{ij}\partial _{s}R_{j}\left( s\right)  \label{19}
\end{eqnarray}%
Both (18) and (19) are standard forumulae for the area $v$ enclosed by the
curve $\vec{R}\left( s\right) $\cite{Area}$.$

\paragraph{Isoperimetric Inequality}

There are numerous proofs of this inequality\cite{Hopf}\cite{Isoperimetric}.
One rather cute and straightforward proof uses Fourier Series. Since $%
\partial V$ is a closed curve (the boundary of a boundary is zero) the
functions $R_{i}\left( s\right) $ are periodic in the length 
\begin{equation}
L=\int\limits_{\partial V}ds  \label{20}
\end{equation}%
and so can be written as $R_{i}\left( s\right) =\sum_{n=-\infty }^{+\infty }%
\tilde{R}_{i,n}\exp \left[ n2\pi is/L\right] $with $\tilde{R}_{i,n}=\tilde{R}%
_{i,-n}^{\ast }$ so that $X_{i}\left( s\right) $ is real ("$\ast $"
indicates complex conjugation). Substituting into (5) and using some very
simple inequalities yields one proof of the isoperimetric inequality (see
for example, Luthy in \cite{Isoperimetric}).

The approach we will use use is to extremize $v$ with the constraint that $%
\left( \partial _{s}\vec{R}\left( s\right) \right) ^{2}=1$ imposed using a
Lagrange multiplier $\lambda $. Minimizing 
\begin{equation}
v=\frac{1}{2}\int\limits_{\partial V}ds\;R_{i}\left( s\right) \varepsilon
_{ij}\partial _{s}R_{j}\left( s\right) +\lambda \left(
\int\limits_{0}^{L}ds\left( \partial _{s}R_{i}\left( s\right) \right)
^{2}-L\right)   \label{21}
\end{equation}%
with respect to $R_{i}\left( s\right) ,~$and $\lambda $ yields%
\begin{eqnarray}
0 &=&\frac{\delta v}{\delta R_{1}\left( s\right) }=\partial _{s}R_{2}\left(
s\right) -2\lambda \partial _{s}^{~2}R_{1}\left( s\right)   \notag \\
0 &=&\frac{\delta v}{\delta R_{2}\left( s\right) }=-\partial _{s}R_{1}\left(
s\right) -2\lambda \partial _{s}^{~2}R_{2}\left( s\right)   \label{22} \\
0 &=&\frac{\partial v}{\partial \lambda }=\int\limits_{0}^{L}ds\left(
\partial _{s}R_{i}\left( s\right) \right) ^{2}-L  \notag
\end{eqnarray}%
Here $\delta /\delta R_{i}\left( s\right) $ indicates functional
differentiation\cite{Zee}. Combining the first and second equations in (22)
gives%
\begin{equation}
\left( \partial _{s}R_{i}\right) +2\lambda \partial _{s}^{~2}\left( \partial
_{s}R_{i}\right) =0\text{ for }i=1,2  \label{23}
\end{equation}%
Taking $\partial _{s}R_{1}\left( s=0\right) =0$ so that $\partial
_{s}R_{2}\left( s=0\right) =1,$ the solutions are 
\begin{eqnarray}
\partial _{s}R_{1}\left( s\right)  &=&-\sin \left( s/2\lambda \right)  
\notag \\
\partial _{s}R_{2}\left( s\right)  &=&\cos \left( s/2\lambda \right) 
\label{24}
\end{eqnarray}%
where the minus sign was chosen for convenience. Integrating both sides
yields%
\begin{eqnarray}
R_{1}\left( s\right)  &=&2\lambda \cos \left( s/2\lambda \right) +R_{0,1} 
\notag \\
R_{2}\left( s\right)  &=&2\lambda \sin \left( s/2\lambda \right) +R_{0,2}
\label{25}
\end{eqnarray}%
which is a circle of radius 2$\lambda $ centered at the arbitrary position $%
\vec{R}_{0}=\left( R_{0,1},R_{0,2}\right) .$ Note that the radius $2\lambda $
is indeterminate since substituting these solutions into the third equation
above yields an identity. This is as it should be since the circle does not
have to be a particular size to maximize an unspecified value of $v.$ It
just has to be a circle. Of course once a particular value of $v$ is given
then $\lambda $ can be determined. Note that the above calculation shows
only that the circle is one shape which extremizes the area. In other words,
we have only shown that the circle is a local extremeum, to\ formally
complete the proof requires showing that\ it is the global extremum.

We will not show it here but isoperimetric inequality is generalizable to
arbitrary dimensions with the result that the shape that maximizes the
"volume" for a given surface "area" is always a "ball"\cite{Hopf}\cite%
{Isoperimetric}, i.e., circle for $D=2,$ sphere for $D=3,$ etc.

The isoperimetric inequality can be used to solve Dido's problem which is to
find the open curve of fixed length $L$ which encloses the maximum area when
the end points are connected by a straight line of arbitrary length\cite%
{Dido}. Combining this closed curve with its reflection relative to the
straight line yields a closed curve of length 2$L.$ Via the isoperimetric
inequality the circle with circumference $2L$ encloses the maximim area and
hence the original open curve which maximizes the area is a semicircle.

\paragraph{Stokes Law}

We now use (5) to derive the planar version of Stokes law as used in
classical electrodynamics\cite{Stokes}. It should come as no surprise that
we can relate Gauss's law to Stokes law and indeed to Green's theorem as
well since they are all different aspects of the general version of Stokes
theorem represented compactly using differential forms\cite{Baez}. Consider
the integral over an area $V$ in the $x_{1}x_{2}$ plane of the curl of a
smooth 3 dimensional vector field $\vec{F}\left( \vec{x}\right) $. Let
indices from the beginning of the aphabet $a,b,c,\cdots $ range over 1,2,3
and indices from the middle of the aphabet $i,j,k,\cdots $ range over 1,2.
This way $\vec{F}\left( \vec{x}\right) =F_{a}\left( \vec{x}\right) \hat{u}%
_{a}$ and $\vec{x}=x_{a}\hat{u}_{a}$ with $\hat{u}_{a}$ the unit vectors
with respect to $x_{a}.$ Represent $\vec{F}\left( \vec{x}\right) $ as a
Fourier transform%
\begin{equation}
F_{i}\left( \vec{x}\right) =\int d^{3}\beta \tilde{F}_{i}\left( \vec{\beta}%
\right) e^{i\vec{\beta}\cdot \vec{x}}  \label{26}
\end{equation}%
with $\vec{\beta}=\beta _{a}\hat{u}_{a}.$ Again we assume that orders of
integration can be freely exchanged. Using index notation the $a^{th}$
component of the curl can be written 
\begin{equation}
\left( \vec{\partial}\times \vec{F}\left( \vec{x}\right) \right)
_{a}=\varepsilon _{abc}\partial _{b}F_{c}\left( \vec{x}\right)  \label{27}
\end{equation}%
Here $\varepsilon _{abc}$ is the $D=3$ totally antisymmetric Levi-Civita
tensor defined as $+$1 for even permutations of $a,b,c=1,2,3$, $-1$ for odd
permutations and 0 if any two of the indices have the same value. Repeated\ $%
a,b,c,\cdots $ indices are summed over the range 1,2,3. The normal to $V,$
which lies in the $x_{1}x_{2}$ plane, is $\hat{u}_{3}$ and so 
\begin{equation}
\int\limits_{V}d^{2}x~\hat{u}_{3}\cdot \left( \vec{\partial}\times \vec{F}%
\left( \vec{x}_{V}\right) \right) =\int\limits_{V}d^{2}x\varepsilon
_{3bc}\partial _{b}F_{c}\left( \vec{x}_{V}\right)  \label{28}
\end{equation}%
where $\vec{x}_{V}=x_{i}\hat{u}_{i}.$ But $\varepsilon _{3bc}$vanishes
unless $b$ and $c$ are restricted to the range 1,2 and so $\varepsilon
_{3bc}\partial _{b}F_{c}\left( \vec{x}\right) =\varepsilon _{ij}\partial
_{i}F_{j}\left( \vec{x}\right) $ where $\varepsilon _{jk}$ are the elements
of the 2$\times $2 Levi-Civita matrix $\varepsilon $ defined above. Putting
this together and using (5) and (26) gives 
\begin{eqnarray}
\int\limits_{V}d^{2}x~\hat{u}_{3}\cdot \left( \vec{\partial}\times \vec{F}%
\left( \vec{x}_{V}\right) \right) &=&\int\limits_{V}d^{2}x~\varepsilon
_{ij}\partial _{i}F_{j}\left( \vec{x}_{V}\right)  \notag \\
&=&\int d^{2}\beta i\beta _{i}\varepsilon _{ij}\tilde{F}_{j}\left( \vec{\beta%
}\right) \int\limits_{V}d^{2}x~e^{i\vec{\beta}\cdot \vec{x}_{V}}  \notag \\
&=&\int d^{2}\beta \beta _{i}\varepsilon _{ij}\tilde{F}_{j}\left( \vec{\beta}%
\right) \frac{\beta _{k}\varepsilon _{kl}}{\beta ^{2}}\int\limits_{\partial
V}ds~t_{l}\left( s\right) e^{i\vec{\beta}\cdot \vec{R}\left( s\right) } 
\notag \\
&=&\int\limits_{\partial V}ds~t_{i}\left( s\right) \int d^{2}\beta e^{i\vec{%
\beta}\cdot \vec{R}\left( s\right) }\tilde{F}_{i}\left( \vec{\beta}\right) 
\notag \\
&=&\int\limits_{\partial V}ds~\hat{t}\left( s\right) \cdot \vec{F}\left( 
\vec{R}\left( s\right) \right)  \label{29}
\end{eqnarray}%
which is the planar form of Stokes law. To get the fourth line we used%
\begin{equation}
\beta _{1}\int\limits_{\partial V}ds~t_{1}\left( s\right) e^{i\vec{\beta}%
\cdot \vec{R}\left( s\right) }=-\beta _{2}\int\limits_{\partial
V}ds~t_{2}\left( s\right) e^{i\vec{\beta}\cdot \vec{R}\left( s\right) }
\label{30}
\end{equation}%
which follows from 
\begin{eqnarray}
\int\limits_{\partial V}ds~i\vec{\beta}\cdot \hat{t}\left( s\right) e^{i\vec{%
\beta}\cdot \vec{R}\left( s\right) } &=&\int\limits_{\partial V}ds~i\vec{%
\beta}\cdot \partial _{s}\vec{R}\left( s\right) e^{i\vec{\beta}\cdot \vec{R}%
\left( s\right) }  \notag \\
&=&\int\limits_{\partial V}ds~\partial _{s}e^{i\vec{\beta}\cdot \vec{R}%
\left( s\right) }  \notag \\
&=&0  \label{31}
\end{eqnarray}

\subsection{Polygons}

A polygon with $N$ sides and $N$ vertices in two dimensions can be defined
by its vertices, arranged in a particular order, $\vec{v}_{1},\vec{v}%
_{2},\cdots ,\vec{v}_{N}$ with $\vec{v}_{i}=\left( x_{1i},x_{2i}\right) .$
For computational convenience it is useful to let $\vec{v}_{N+1}=\vec{v}_{1}$%
. We consider only orientable polygons, i.e., none of the edges cross over
or intersect one another. The integral over $\partial V$ in (7) can be
evaluated explicitly in this case with the result%
\begin{eqnarray}
\phi \left( \vec{\beta}\right) &=&\frac{\vec{\beta}}{i\beta ^{2}}\cdot
\int\limits_{\partial V}d^{D-1}s\;\hat{n}\left( \vec{s}\right) \cdot e^{i%
\vec{\beta}\cdot \vec{R}\left( s\right) }  \notag \\
&=&-\frac{1}{\beta ^{2}}\sum_{n=1}^{N}\frac{\vec{\beta}_{\bot }\cdot \left( 
\vec{v}_{n+1}-\vec{v}_{n}\right) }{\vec{\beta}\cdot \left( \vec{v}_{n+1}-%
\vec{v}_{n}\right) }\left( \exp \left( i\vec{\beta}\cdot \vec{v}%
_{n+1}\right) -\exp \left( i\vec{\beta}\cdot \vec{v}_{n}\right) \right)
\label{32}
\end{eqnarray}%
where 
\begin{equation}
\vec{\beta}_{\bot }\equiv \vec{\beta}\cdot \varepsilon =\left( \beta
_{1},\beta _{2}\right) \cdot 
\begin{bmatrix}
0 & 1 \\ 
-1 & 0%
\end{bmatrix}%
=\left( \beta _{2},-\beta _{1}\right)  \label{33}
\end{equation}%
The "$\cdot $" stands for standard matrix vector multiplication.

Equating the surface integral to the series expansion in terms of moments
gives%
\begin{eqnarray}
&&\sum_{n=0}^{\infty }\frac{i^{n}}{n!}\sum_{p=0}^{n}\frac{n!}{p!\left(
n-p\right) !}\beta _{1}^{p}\beta _{2}^{\left( n-p\right) }M\left(
x_{1}^{p}x_{2}^{n-p}\right)  \notag \\
&=&-\frac{1}{\beta ^{2}}\sum_{n=1}^{N}\frac{\vec{\beta}_{\bot }\cdot \left( 
\vec{v}_{n+1}-\vec{v}_{n}\right) }{\vec{\beta}\cdot \left( \vec{v}_{n+1}-%
\vec{v}_{n}\right) }\left( \exp \left( i\vec{\beta}\cdot \vec{v}%
_{n+1}\right) -\exp \left( i\vec{\beta}\cdot \vec{v}_{n}\right) \right)
\label{34}
\end{eqnarray}%
Now expand the right hand side in powers of $\beta _{i}$

\begin{align}
& \frac{1}{\beta ^{2}}\sum_{n=1}^{N}\frac{\vec{\beta}_{\bot }\cdot \left( 
\vec{v}_{n+1}-\vec{v}_{n}\right) }{\vec{\beta}\cdot \left( \vec{v}_{n+1}-%
\vec{v}_{n}\right) }\left( \exp \left( i\vec{\beta}\cdot \vec{v}%
_{n+1}\right) -\exp \left( i\vec{\beta}\cdot \vec{v}_{n}\right) \right) 
\notag \\
& =\frac{1}{\beta ^{2}}\sum_{n=1}^{N}\frac{\vec{\beta}_{\bot }\cdot \left( 
\vec{v}_{n+1}-\vec{v}_{n}\right) }{\vec{\beta}\cdot \left( \vec{v}_{n+1}-%
\vec{v}_{n}\right) }\left( \sum_{m=1}^{\infty }\frac{\left( i\vec{\beta}%
\cdot \vec{v}_{n+1}\right) ^{m}-\left( i\vec{\beta}\cdot \vec{v}_{n}\right)
^{m}}{m!}\right)  \notag \\
& =\frac{1}{\beta ^{2}}\sum_{m=1}^{\infty }\sum_{n=1}^{N}\frac{i}{m!}\vec{%
\beta}_{\bot }\cdot \left( \vec{v}_{n+1}-\vec{v}_{n}\right)
\sum_{p=0}^{m-1}\left( i\vec{\beta}\cdot \vec{v}_{n+1}\right) ^{m-1-p}\left(
i\vec{\beta}\cdot \vec{v}_{n}\right) ^{p}  \label{35}
\end{align}%
where $\beta ^{2}=\vec{\beta}^{2}=\beta _{i}\beta _{i}=\beta _{1}^{~2}+\beta
_{2}^{~2}.$ In the second step we have used the identity%
\begin{equation}
a^{n}-b^{n}=\left( a-b\right) \sum_{m=0}^{n-1}a^{n-1-m}b^{m}  \label{36}
\end{equation}%
Substituting (35) into (34) gives%
\begin{eqnarray}
&&\sum_{n=0}^{\infty }\frac{i^{n}}{n!}\sum_{p=0}^{n}\frac{n!}{p!\left(
n-p\right) !}\beta _{1}^{p}\beta _{2}^{n-p}M\left(
x_{1}^{p}x_{2}^{n-p}\right)  \notag \\
&=&-\frac{1}{\beta ^{2}}\sum_{m=1}^{\infty }\sum_{n=1}^{N}\frac{i}{m!}\vec{%
\beta}_{\bot }\cdot \left( \vec{v}_{n+1}-\vec{v}_{n}\right)
\sum_{p=0}^{m-1}\left( i\vec{\beta}\cdot \vec{v}_{n+1}\right) ^{m-1-p}\left(
i\vec{\beta}\cdot \vec{v}_{n}\right) ^{p}  \label{37}
\end{eqnarray}%
Letting $\vec{\beta}=\beta \hat{\beta}$ and $\vec{\beta}_{\bot }=\beta \hat{%
\beta}_{\bot }$ we see that the $n^{th}$ term on the left hand side is
proportional to $\beta ^{n}$ whereas the $m^{th}$ on the right is
proportional to $\beta ^{m-2}.$ Hence considering each side as a power
series in $\beta $ the coefficients of the $m$ term on the right must equal
the coefficient of $n=m-2$ term on the left. Below we write out the general
relation between $M\left( x_{1}^{a}x_{2}^{b}\right) $ and powers of the
vertices. Here we begin by considering the first few terms individually.

\subsubsection{$m=1$ term}

The $m=1$ term on the right has no corresponding term on the left of (34)
and so we must have

\begin{equation}
0=\frac{i}{\beta ^{2}}\vec{\beta}_{\bot }\cdot \sum_{n=1}^{N}\left( \vec{v}%
_{n+1}-\vec{v}_{n}\right)  \label{38}
\end{equation}%
The fact that this should vanish follows from the fact that it scales as $%
1/\beta $ whereas the left hand side remains finite as $\beta \rightarrow 0$%
. Also this term is imaginary and the result must be real. And indeed this
term does vanish, trivially, since we have defined $\vec{v}_{N+1}=\vec{v}%
_{1} $ and the sum of the directed sides, $\vec{v}_{n+1}-\vec{v}_{n}$, of a
closed polygon must vanish. For the case where the boundary $\partial V$ is
everywhere smooth the form of the dominant term as $\beta \rightarrow 0$ is
given by 
\begin{equation}
\underset{\beta \rightarrow 0}{\lim }\frac{\vec{\beta}}{i\beta ^{2}}\cdot
\int\limits_{\partial V}ds\;\hat{n}\left( s\right) e^{i\vec{\beta}\cdot \vec{%
R}\left( s\right) }\rightarrow \frac{\hat{\beta}}{i\beta }\cdot
\int\limits_{\partial V}ds\;\hat{n}\left( s\right) \Rightarrow
\int\limits_{\partial V}ds\;\hat{n}\left( s\right) =0  \label{39}
\end{equation}%
The last equality follows from the fact that this term must vanish for any $%
\left( \text{non-zero}\right) $ $\vec{\beta}$. The results in $\left( \
\right) $ and $\left( \ \right) $ can be seen as a simple version of the
more formal statement that the boundary of a boundary is zero\cite{Boundary}%
, i.e., the curve which bounds an area in two dimensions has no end points.

\paragraph{Hopfs Umlaufsatz}

The results in (38) and (39) can also be related to Hopfs Umlaufsatz\cite%
{Hopf} which states that the tangent vector (or the normal vector or a
linear combination thereof) to a smooth or piecewise smooth orientable curve
in two dimensions rotates by $2\pi $ on making a complete circuit of the
curve if the curve is traversed in a counterclockwise direction and by $%
-2\pi $ if traversed in a clockwise direction. For nonorientable curves,
i.e., curves which crossover or intesect themselves it is an integer
multiple of $2\pi $ with the integer value counting the\ net number of
signed (+/$-$ for counterclockwise/clockwise) loops. This seems obvious but
in fact is rather subtle to prove\cite{Hopf}.

The following is an argument in favor of the Hopf Umlaufsatz but is not a
proof in the strict mathematical sense. Consider first the piecewise smooth,
actually piecewise linear, case of a polygon defined by the vertices $\vec{v}%
_{1},\cdots ,\vec{v}_{N}$ \ with $\vec{v}_{N+1}=\vec{v}_{1}.$ Let $\vec{l}%
_{n}=\vec{v}_{n+1}-\vec{v}_{n}.$which are the directed sides of the polygon
taken in order stepping around the polygon. We can write $\vec{l}%
_{n+1}=\lambda _{n+1}R\left( \theta _{n+1}\right) \cdot \vec{l}_{n}$ for $%
n=1,2,\cdots ,N+1$with $\lambda _{n}>0$ and $R\left( \theta \right) $ the
two dimensional rotation matrix%
\begin{equation}
R\left( \theta \right) =\left[ 
\begin{array}{cc}
\cos \left[ \theta \right] & \sin \left[ \theta \right] \\ 
-\sin \left[ \theta \right] & \cos \left[ \theta \right]%
\end{array}%
\right]  \label{40}
\end{equation}%
Thus $\vec{l}_{n+1}=\lambda _{n+1}R\left( \theta _{n+1}\right) \cdot \vec{l}%
_{n}$ defines each successive side vector $\vec{l}_{n}$ as a rotated and
rescaled copy of the previous side vector. Combining these relations for the
entire polygon gives 
\begin{align}
\vec{l}_{N+1}& =\lambda _{N+1}\lambda _{N}\cdots \lambda _{2}R\left( \theta
_{N+1}\right) R\left( \theta _{N}\right) \cdots R\left( \theta _{2}\right) 
\vec{l}_{1}  \notag \\
& =\lambda _{N+1}\lambda _{N}\cdots \lambda _{2}R\left( \theta _{N+1}+\theta
_{N}+\cdots +\theta _{2}\right) \vec{l}_{1}  \notag \\
& =\vec{l}_{1}  \label{41}
\end{align}%
which demands that 
\begin{equation}
\lambda _{N+1}\lambda _{N}\cdots \lambda _{2}R\left( \theta _{N+1}+\theta
_{N}+\cdots +\theta _{2}\right) =1  \label{42}
\end{equation}%
and so $\lambda _{N+1}\lambda _{N}\cdots \lambda _{2}=1$ and $\theta
_{N+1}+\theta _{N}+\cdots +\theta _{2}=$ integer $\times \left( 2\pi \right)
.$ This argument does not by itself specify the integer value as being
unity. The fact that the integer needs to be 1 follows from the requirement
of the polygon being orientable.

The same type of argument can be applied to the smooth curve case. Write the
tangent vector $\hat{t}$ at distance $s$ along the curve as%
\begin{equation}
\hat{t}\left( s+ds\right) =R\left( \partial _{s}\theta \left( s\right)
ds\right) \cdot \hat{t}\left( s\right)  \label{43}
\end{equation}%
where $\theta \left( s\right) $ is the angle between the tangent vector at $%
s $ and the tangent vector at $s=0$ so that%
\begin{equation}
\hat{t}\left( s\right) =R\left( \theta \left( s\right) \right) \cdot \hat{t}%
\left( 0\right)  \label{44}
\end{equation}%
Then using the fact that the curve is smooth ($\partial _{s}\hat{t}\left(
s\right) $ is everwhere defined) and closed we have, with $L$ the total
distance around the curve, that%
\begin{align}
\hat{t}\left( L\right) & =R\left( \theta \left( L\right) \right) \cdot \hat{t%
}\left( 0\right)  \notag \\
& =\hat{t}\left( 0\right)  \label{45}
\end{align}%
and so $\theta \left( L\right) $ must be an integer times $2\pi .$ Again
this result must be combined with the fact that the curve is orientable to
specify the integer value as unity.

\subsubsection{$m=2$ term}

The $m=2$ term on the right equals the $n=0$ term on the left and so we have

\begin{align}
M\left( 1\right) & =-\frac{i}{2}\frac{1}{\beta ^{2}}\sum_{n=1}^{N}\vec{\beta}%
_{\bot }\cdot \left( \vec{v}_{n+1}-\vec{v}_{n}\right) \sum_{p=0}^{1}\left( i%
\vec{\beta}\cdot \vec{v}_{n+,1}\right) ^{1-p}\left( i\vec{\beta}\cdot \vec{v}%
_{n}\right) ^{p}  \notag \\
& =-\frac{i}{2}\frac{1}{\beta ^{2}}\sum_{n=1}^{N}\left( \vec{\beta}_{\bot
}\cdot \left( \vec{v}_{n+1}-\vec{v}_{n}\right) \right) \left( i\vec{\beta}%
\cdot \vec{v}_{n+1}+i\vec{\beta}\cdot \vec{v}_{n}\right)  \notag \\
& =\frac{1}{2}\sum_{n=1}^{N}\left( \hat{\beta}_{\bot }\cdot \left( \vec{v}%
_{n+1}-\vec{v}_{n}\right) \right) \left( \hat{\beta}\cdot \left( \vec{v}%
_{n+1}+\vec{v}_{n}\right) \right)  \label{46}
\end{align}%
But by definition $M\left( 1\right) $ is the area of the polygon. Defining $%
\vec{l}_{n}=\vec{v}_{n+1}-\vec{v}_{n}$ = the vector from vertex $n$ to
vertex $n+1$ and $\vec{c}_{n}=\left( \vec{v}_{n+1}+\vec{v}_{n}\right) /2=$
position of the center of side $n$ we have 
\begin{equation}
M\left( 1\right) =Area=\sum_{n=1}^{N}\left( \hat{\beta}_{\bot }\cdot \vec{l}%
_{n}\right) \left( \hat{\beta}\cdot \vec{c}_{n}\right)  \label{47}
\end{equation}%
Taking $\hat{\beta}=\left( 0,1\right) $ gives $\hat{\beta}_{\bot }=\left(
-1,0\right) $ and so%
\begin{equation}
Area=-\sum_{n=1}^{N}l_{n,1}c_{n,2}  \label{48}
\end{equation}%
The geometric interpretation of this is straightforward. Each of the $%
l_{n,1}c_{n,2}$ terms correspond to the area of a 4-sided polygon of width $%
\left\vert l_{n,1}\right\vert $ in the $x_{1}$ direction and mean height $%
\left\vert c_{n,2}\right\vert $ in the $x_{2}$ direction. If we take all the
vertices to lie in the first quadrant, $v_{n,i}>0$ for $i=1,2$ then all the $%
c_{n,2}$ are positive but the $l_{n,1}$ change sign depending on whether $%
\vec{l}_{n}$ points generally in the $+x_{1}$ or $-$ $x_{1}$ direction. If
we consider that the $n=1,\cdots ,N$ ordering corresponds to following the
vertices in a counterclockwise direction around the polygon then the
positive $l_{n,1}$ will generally run along the bottom sides of the net
polygon and the negative $l_{n,1}$ will generally run along the top sides of
the net polygon, then the area of the net polygon is the total area of all
the 4-sided polygons along the bottom of the net polygon subtracted from the
area of the 4-sided polygons along the top of the net polygon.

Taking 1/2 the sum of (47) with $\hat{\beta}=\left( 0,1\right) $ and with $%
\hat{\beta}=\left( 1,0\right) $ we find that the area can also be written as 
$\frac{1}{2}\sum_{n=1}^{N}\det \left[ M_{n}\right] $ where the elements of
the 2$\times 2$ matriices $M_{n}$ are given by $M_{n,ij}=v_{n,i}v_{n+1,j},$%
which is the standard result\cite{PolygonArea}.

\paragraph{Area Element}

We now show that the area element $dv$ of a 2 dimensional surface $\partial
V $ defined by $\vec{R}\left( \vec{s}\right) $, which is a vector in 3
dimensional Euclidean space and $\vec{s}=\left( s_{1},s_{2}\right) $ labels
points on $\partial V,$ is given by 
\begin{equation}
dv=\sqrt{\left\vert \det \left[ \partial _{s_{i}}\vec{R}\left( \vec{s}%
\right) \cdot \partial _{s_{j}}\vec{R}\left( \vec{s}\right) \right]
\right\vert }ds_{1}ds_{2}=\sqrt{\left\vert \det \left[ \hat{t}_{i}\left( 
\vec{s}\right) \cdot \hat{t}_{j}\left( \vec{s}\right) \right] \right\vert }%
ds_{1}ds_{2}  \label{49}
\end{equation}%
The two vectors $d\vec{R}_{i}=\partial _{s_{i}}\vec{R}\left( \vec{s}\right)
ds_{i}=\hat{t}_{i}\left( \vec{s}\right) ds_{i}$ for $i=1,2$ (no sum on $i$)
form a parellelogram with area $dv$ whose vertices are $\vec{v}_{1}=\left(
0,0\right) ,$ $\vec{v}_{2}=d\vec{R}_{1},$ $\vec{v}_{3}=d\vec{R}_{1}+d\vec{R}%
_{2},$ $\vec{v}_{4}=d\vec{R}_{2}$ and $\vec{v}_{5}=\vec{v}_{1}$ all of which
lie on $\partial V.$ (For simplicity we have assumed that $s_{i}$ has units
of length so that $\partial _{s_{i}}\vec{R}\left( \vec{s}\right) =\hat{t}%
_{i}\left( \vec{s}\right) $ with $\hat{t}_{i}\left( \vec{s}\right) $
automatically normalized to unity.) The area of this parallelogram $dv$ can
be computed directly from (46). 
\begin{eqnarray}
dv &=&\frac{1}{2}\left[ 
\begin{array}{c}
\left( \hat{\beta}_{\bot }\cdot d\vec{R}_{1}\right) \left( \hat{\beta}\cdot d%
\vec{R}_{1}\right) +\left( \hat{\beta}_{\bot }\cdot d\vec{R}_{2}\right)
\left( \hat{\beta}\cdot \left( 2d\vec{R}_{1}+d\vec{R}_{2}\right) \right) \\ 
+\left( \hat{\beta}_{\bot }\cdot -d\vec{R}_{1}\right) \left( \hat{\beta}%
\cdot \left( d\vec{R}_{1}+2d\vec{R}_{2}\right) \right) +\left( \hat{\beta}%
_{\bot }\cdot -d\vec{R}_{2}\right) \left( \hat{\beta}\cdot d\vec{R}%
_{2}\right)%
\end{array}%
\right]  \notag \\
&=&\left( \hat{\beta}\cdot \varepsilon \cdot d\vec{R}_{2}\right) \left( \hat{%
\beta}\cdot d\vec{R}_{1}\right) -\left( \hat{\beta}\cdot \varepsilon \cdot d%
\vec{R}_{1}\right) \left( \hat{\beta}\cdot d\vec{R}_{2}\right)  \notag \\
&=&\left( \left( \hat{\beta}\cdot \varepsilon \cdot \hat{t}_{2}\right)
\left( \hat{\beta}\cdot \hat{t}_{1}\right) -\left( \hat{\beta}\cdot
\varepsilon \cdot \hat{t}_{1}\right) \left( \hat{\beta}\cdot \hat{t}%
_{2}\right) \right) ds_{1}ds_{2}  \label{50}
\end{eqnarray}%
The vector $\hat{\beta}$ can be written as an arbitrary linear combination
of the unit vectors $\hat{t}_{i}\left( \vec{s}\right) $. Letting $\hat{\beta}%
=\hat{t}_{1}$ gives%
\begin{eqnarray}
dv &=&\hat{t}_{1}\cdot \varepsilon \cdot \hat{t}_{2}ds_{1}ds_{2}  \notag \\
&=&\left( t_{1,1}t_{2,2}-t_{1,2}t_{2,1}\right) ds_{1}ds_{2}  \label{51}
\end{eqnarray}%
where $t_{i,a}$ are the components of $\hat{t}_{i}$ in an orthonormal
coordinate system indexed by $a=1,2$ erected on $\partial V$ at $\vec{s}.$
But%
\begin{eqnarray}
\sqrt{\left\vert \det \left[ \hat{t}_{i}\cdot \hat{t}_{j}\right] \right\vert 
}ds_{1}ds_{2} &=&\sqrt{\left\vert \det 
\begin{bmatrix}
\hat{t}_{1}\cdot \hat{t}_{1} & \hat{t}_{1}\cdot \hat{t}_{2} \\ 
\hat{t}_{2}\cdot \hat{t}_{1} & \hat{t}_{2}\cdot \hat{t}_{2}%
\end{bmatrix}%
\right\vert }ds_{1}ds_{2}  \notag \\
&=&\sqrt{\left\vert \left( \hat{t}_{1}\cdot \hat{t}_{1}\right) \left( \hat{t}%
_{2}\cdot \hat{t}_{2}\right) -\left( \hat{t}_{1}\cdot \hat{t}_{2}\right)
^{2}\right\vert }ds_{1}ds_{2}  \notag \\
&=&\left( t_{1,1}t_{2,2}-t_{1,2}t_{2,1}\right) ds_{1}ds_{2}  \notag \\
&=&dv  \label{52}
\end{eqnarray}%
For the purposes of the above derivation even though $\hat{t}_{1}\cdot \hat{t%
}_{1}=\hat{t}_{2}\cdot \hat{t}_{2}=1$ it is more convenient leave them as $%
\hat{t}_{1}\cdot \hat{t}_{1}$ and $\hat{t}_{2}\cdot \hat{t}_{2}.$ Finally
note that from simple geometry we have $dv=\left\vert d\vec{R}%
_{1}\right\vert \left\vert d\vec{R}_{2}\right\vert \sin \left( \theta
\right) $ where $\theta $ is the angle between $d\vec{R}_{1}$ and $d\vec{R}%
_{2}$, and if we do use $\hat{t}_{1}\cdot \hat{t}_{1}=\hat{t}_{2}\cdot \hat{t%
}_{2}=1$ then $\sqrt{\left\vert \det \left[ \hat{t}_{i}\cdot \hat{t}_{j}%
\right] \right\vert }=\sqrt{1-\left( \hat{t}_{1}\cdot \hat{t}_{2}\right) ^{2}%
}=\sqrt{1-\cos \left( \theta \right) ^{2}}=\sin \left( \theta \right) $ and
we get the same result for $dv$. In general it is much more convenient to
work with $\sqrt{\left\vert \det \left[ \hat{t}_{i}\cdot \hat{t}_{j}\right]
\right\vert }$ than other forms since the inner product $\hat{t}_{i}\cdot 
\hat{t}_{j}$ is coordinate independent.

\subsubsection{$m=3$ term}

The $m=3$ term on the right hand side corresponds to the $n=1$ term on the
left and so after cancelling $i$ from both sides we have

\begin{equation}
\hat{\beta}_{1}M\left( x_{1}\right) +\hat{\beta}_{2}M\left( x_{2}\right) =%
\frac{1}{3!}\sum_{n=1}^{N}\hat{\beta}_{\bot }\cdot \left( \vec{v}_{n+1}-\vec{%
v}_{n}\right) \left( 
\begin{array}{c}
\left( \hat{\beta}\cdot \vec{v}_{n+1}\right) ^{2} \\ 
+\left( \hat{\beta}\cdot \vec{v}_{n+1}\right) \left( \hat{\beta}\cdot \vec{v}%
_{n}\right) \\ 
+\left( \hat{\beta}\cdot \vec{v}_{n}\right) ^{2}%
\end{array}%
\right)  \label{53}
\end{equation}%
It follows from the definition of $M$ that%
\begin{equation}
M\left( x_{i}\right) =Area\times X_{i}  \label{54}
\end{equation}%
where $X_{i}$ is the "center of mass" or centroid of the polygon in the $%
i=1,2$ directions.

For $\hat{\beta}=\left( 1,0\right) $, $\hat{\beta}_{\bot }=\left( 0,1\right) 
$ we have%
\begin{equation}
M\left( x_{1}\right) =\frac{1}{3!}\sum_{n=1}^{N}\left(
v_{n+1,2}-v_{n,2}\right) \left(
v_{n+1,1}^{2}+v_{n,1}^{2}+v_{n+1,1}v_{n,1}\right)  \label{55}
\end{equation}%
and for $\hat{\beta}=\left( 0,1\right) ,$ $\hat{\beta}_{\bot }=\left(
-1,0\right) $ we have%
\begin{equation}
M\left( x_{2}\right) =-\frac{1}{3!}\sum_{n=1}^{N}\left(
v_{n+1,1}-v_{n,1}\right) \left(
v_{n+1,2}^{2}+v_{n,2}^{2}+v_{n+1,2}v_{n,2}\right)  \label{56}
\end{equation}

Explicit evaluation of 
\begin{equation}
\int_{V}dx_{1}dx_{2}x_{i}  \label{57}
\end{equation}%
with $i=1$ or $2$ by substituting $x_{i}=\vec{\partial}\cdot \left( x_{i}^{2}%
\hat{x}_{i}/2\right) ,$ with no sum on $i,$ yields the same result as in
(55) and (56).

\subsection{Moments and Shapes of Polygons}

We begin this section by using a version of (5) modified to live in the
complex plane to provide an alternative derivation of the result of Davis
which is a generalization from triangles to polygons of the
Motzkin-Schoenberg formula\cite{Davis}. This result is also related to the
so-called "shape from moments" problem which is to find the ordered vertices
of a polygon given an appropriate set of the polygon moments\cite{Milanfar}%
\cite{SIAM}. As shown by Milanfar\cite{Milanfar}, in the complex plane with $%
z=x+iy,$ the result of Davis can be written%
\begin{equation}
\int_{V}dxdy\partial _{z}^{~2}h\left( z\right) =\frac{i}{2}%
\sum_{n=1}^{N}\left( \frac{z_{n-1}^{\ast }-z_{n}^{\ast }}{z_{n-1}-z_{n}}-%
\frac{z_{n}^{\ast }-z_{n+1}^{\ast }}{z_{n}-z_{n+1}}\right) h\left(
z_{n}\right)  \label{58}
\end{equation}%
Here $V$ is a simply connected orientable polygon with vertices $%
z_{n}=x_{n}+iy_{n}$ $\left( z_{n}^{\ast }=z_{n}-iy_{n}\right) $, $n=1,\cdots
,N$, in the complex plane and the function $h\left( z\right) $ is analytic
(= holomorphic = regular) in the closure of $V$. In the sum\ we have let $%
z_{0}=z_{N}$ and $z_{1}=z_{N+1}.$ For the remainder of this section we
follow the standard notation for variables in the complex plane, i.e., we
replace $x_{1}$with $x$, $x_{2}$ with $y,$ etc.$.$

To write $h\left( z\right) $ as a Fourier transform start with the
definition of an analytic function, i.e., that it can be written as a power
series in non-negative powers of $z$%
\begin{equation}
h\left( z\right) =\sum_{n=0}^{\infty }a_{n}z^{n}  \label{59}
\end{equation}%
Now define the function $\tilde{h}\left( \beta \right) $, with $\beta $
real, by 
\begin{equation}
a_{n}=\int d\beta \tilde{h}\left( \beta \right) \frac{\left( i\beta \right)
^{n}}{n!}  \label{60}
\end{equation}%
This may seem restrictive but if we take $\tilde{h}\left( \beta \right) $ to
vanish outside $\left\vert \beta \right\vert \leq 1$ then we can represent $%
\tilde{h}\left( \beta \right) $ as 
\begin{equation}
\tilde{h}\left( \beta \right) =\sum_{n=0}^{\infty }A_{n}P_{n}\left( \beta
\right)  \label{61}
\end{equation}%
where $P_{n}\left( \beta \right) $ are Legendre polynomials. With this
representation the complex coefficients $A_{n}$ can be chosen to satisfy $%
\left( 35\right) $. Again assuming we can interchange sums and integrals at
will we can write 
\begin{equation}
h\left( z\right) =\int d\beta \tilde{h}\left( \beta \right)
\sum_{n=0}^{\infty }\frac{\left( i\beta z\right) ^{n}}{n!}=\int d\beta 
\tilde{h}\left( \beta \right) e^{i\beta z}  \label{62}
\end{equation}%
Using the obvious notation $\left( a,b\right) \cdot \left( x,y\right)
=ax+by, $ we now have%
\begin{align}
\int_{V}dxdy\partial _{z}^{~2}h\left( z\right) & =-\int d\beta \tilde{h}%
\left( \beta \right) \int_{V}dxdy\beta ^{2}e^{i\beta z}  \notag \\
& =\frac{i}{2}\int d\beta \tilde{h}\left( \beta \right) \int_{V}dxdy\left(
\partial _{x},\partial _{y}\right) \cdot \beta \left( 1,-i\right) \exp \left[
i\beta \left( 1,i\right) \cdot \left( x,y\right) \right]  \notag \\
& =\frac{1}{2}\int d\beta \tilde{h}\left( \beta \right) \sum_{n=1}^{N}\frac{%
\beta \left( 1,-i\right) \cdot \varepsilon \cdot \left( \left( x,y\right)
_{n+1}-\left( x,y\right) _{n}\right) }{\beta \left( 1,i\right) \cdot \left(
\left( x,y\right) _{n+1}-\left( x,y\right) _{n}\right) }  \notag \\
& \ \ \ \ \ \ \ \ \ \ \ \ \ \ \ \ \ \ \times \left( \exp \left( i\beta
\left( 1,i\right) \cdot \left( x,y\right) _{n+1}\right) -\exp \left( i\beta
\left( 1,i\right) \cdot \left( x,y\right) _{n}\right) \right)  \notag \\
& =\frac{i}{2}\int d\beta \tilde{h}\left( \beta \right) \sum_{n=1}^{N}\left( 
\frac{z_{n}^{\ast }-z_{n-1}^{\ast }}{z_{n}-z_{n-1}}-\frac{z_{n+1}^{\ast
}-z_{n}^{\ast }}{z_{n+1}-z_{n}}\right) \exp \left( i\beta z_{n}\right) 
\notag \\
& =\frac{i}{2}\sum_{n=1}^{N}\left( \frac{z_{n-1}^{\ast }-z_{n}^{\ast }}{%
z_{n-1}-z_{n}}-\frac{z_{n}^{\ast }-z_{n+1}^{\ast }}{z_{n}-z_{n+1}}\right)
h\left( z_{j}\right)  \label{63}
\end{align}%
In the third line $\varepsilon $ is the Levi-Civita matrix defined in (14).
In the second step we have used (5) and in the last step the definition of $%
h\left( z\right) $ in terms of its Fourier transform. The $\beta $'s cancel
in the coefficient of the exponents in line three and hence we reproduce the
result of Davis that the integral over a polygon of $\partial
_{z}^{~2}h\left( z\right) $ is the sum of $h\left( z\right) $ evaluated at
the vertices of the polygon times coefficients which depend only on the
vertices and not on $h\left( z\right) .$

The moments of a polygon obviously contain the polygon shape information.
The order of the vertices is important as reordering them nominally leads to
a polygon with a different shape and/or can make it nonorientable. For $N$
vertices there are $2N$ independent real numbers corresponding to the $\vec{v%
}_{1},\cdots ,\vec{v}_{N}$ vertices which define the polygon. Thus the
infiinite set of all possible moments of the polygon must be highly
redunant.We will not solve this "shape from moments" problem here. Milanfar
and others \cite{Milanfar}\cite{SIAM} have shown how to solve for the
vertices given a particular set of complex moments which can be easilly
computed from (63) by letting $h\left( z\right) =z^{k}$.Here we merely
present the complete set of relations between all possible moments $M\left(
x^{a},y^{b}\right) $ and the vertices $\left( x,y\right) _{n}$ of a polygon
that follows from (34).

To derive an explicit relation between the moments and the vertices of a an
arbitrary orientable polygon multiply (32) through by $\beta ^{2}=\beta
_{1}^{2}+\beta _{2}^{2}$ and use the fact that the $m=1$ term on the right
hand side vanishes identically. The derivation is facilitated by defining a
function $\theta \left( \cdots \right) $ which vanishes if any one or more
of its arguments is negative and equals 1 otherwise. This function can be
used to keep track of the limits on the sums when the summation indices are
redefined so that the powers of $\beta _{1}$ and $\beta _{2}$ are written as 
$\beta _{1}^{~a}\beta _{2}^{~b}$ on both sides of the equation. Then using
the fact that the coefficient of $\beta _{1}^{~a}\beta _{2}^{~b}$ on the
left must equal the coefficient of $\beta _{1}^{~a}\beta _{2}^{~b}$ on the
right for the same given non-negative integer values of $a$ and $b$ we find,
with $\vec{v}_{n}=\left( x_{1,n},x_{2,n}\right) $%
\begin{align}
& \theta \left( a-2\right) \frac{M\left( x_{1}^{~a-2}x_{2}^{~b}\right) }{%
\left( a-2\right) !b!}+\theta \left( b-2\right) \frac{M\left(
x_{1}^{~a}y_{2}^{~b-2}\right) }{a!\left( b-2\right) !}  \notag \\
& =\theta \left( b-1,a+b-2\right) \sum_{q=0}^{a}\sum_{p=0}^{a+b-1-q}\left( 
\begin{array}{c}
\theta \left( p+q-a\right) \frac{1}{\left( a+b\right) !}\frac{\left(
a+b-1-p\right) !}{q!\left( a+b-1-p-q\right) !}\frac{p!}{\left( a-q\right)
!\left( p+q-a\right) !} \\ 
\times \sum_{n=1}^{N}\left( x_{1,n+1}-x_{1,n}\right)
x_{1,n+1}^{~q}x_{2,n+1}^{~a+b-1-p-q}x_{1,n}^{~a-q}x_{2,n}^{~p+q-a}%
\end{array}%
\right)  \notag \\
& -\theta \left( a-1,a+b-2\right) \sum_{q=0}^{a-1}\sum_{p=0}^{a+b-1-q}\left( 
\begin{array}{c}
\theta \left( p+q+1-a\right) \frac{1}{\left( a+b\right) !}\frac{\left(
a+b-1-p\right) !}{q!\left( a+b-1-p-q\right) !}\frac{p!}{\left( a-q-1\right)
!\left( p+q+1-a\right) !} \\ 
\times \sum_{n=1}^{N}\left( x_{2,n+1}-x_{2,n}\right)
x_{2,n+1}^{~q}x_{2,n+1}^{~a+b-1-p-q}x_{1,n}^{~a-q-1}x_{2,n}^{~p+q+1-a}%
\end{array}%
\right)  \label{64}
\end{align}%
The derivation is straightforward but tedious.

\subsection{Fraunhofer Diffraction}

Fraunhofer diffraction occurs when the light diffracted from an aperture or
opening in an opaque screens is observed in a plane far from aperture itself%
\cite{FourierOptics}. By "far" we mean that the distance between the opaque
screen and the plane of observation, which is by convention taken to be
parallel to the screen, is much larger than the maximum dimension of the
aperture. For the case where the opaque screen lies in the $x_{1}x_{2}$
plane and the illumination is a unit amplitude plane wave of wavelength $%
\lambda $ incident on the screen from the side opposite the observation
plane, the amplitude of the diffracted light at position $x_{i}$ in the
observation plane a distance $L$ away from the screen is given by 
\begin{equation}
A\left( \vec{x}\right) =\int d^{2}x^{\prime }\theta _{V}\left( \vec{x}%
^{\prime }\right) \exp \left[ i\frac{k}{L}x_{i}x_{i}^{\prime }\right]
=\int\limits_{V}d^{2}x^{\prime }\exp \left[ i\frac{k}{L}x_{i}x_{i}^{\prime }%
\right]  \label{65}
\end{equation}%
Here $\theta _{V}\left( x,y\right) =1$ inside the aperture and 0 outside
describes the aperture shape and $k=2\pi /\lambda $. This is essentially the
left hand side of (5) but with 
\begin{equation}
\beta _{i}=\frac{kx_{i}}{L}  \label{66}
\end{equation}%
and without the normalization factor. The intensity of the diffraction
pattern is given by $I\left( \vec{x}\right) =\left\vert A\left( \vec{x}%
\right) \right\vert ^{2}.$

We consider two cases, a circular aperture whose solution follows almost
trivially from (5) and a slit whose solution is already implicit in (32)$.$

For a circular aperture of radius $R$ centered at $x_{i}=0$ it follows from
(5) that 
\begin{equation}
A\left( \vec{x}\right) =\frac{\vec{\beta}R}{i\beta ^{2}}\cdot
\int\limits_{0}^{2\pi }d\varphi \hat{r}\left( \varphi \right) \exp \left[ i%
\vec{\beta}\cdot R\hat{r}\left( \varphi \right) \right]  \label{67}
\end{equation}%
with $\hat{r}\left( \varphi \right) =\vec{x}^{\prime }/\left\vert \vec{x}%
^{\prime }\right\vert .$ Writing $\vec{\beta}\cdot \hat{r}\left( \varphi
\right) =\beta \cos \left( \varphi -\varphi _{\beta }\right) $ where $%
\varphi _{\beta }$ is the angle $\vec{\beta}$ makes with respect to the $%
x_{1}$-axis and $\beta =\left\vert \vec{\beta}\right\vert =k\sqrt{x^{2}+y^{2}%
}/L$, (67) becomes%
\begin{eqnarray}
A\left( \vec{x}\right) &=&\frac{R}{i\beta }\int\limits_{0}^{2\pi }d\varphi
\cos \left( \varphi -\varphi _{\beta }\right) \exp \left[ i\beta R\cos
\left( \varphi -\varphi _{\beta }\right) \right]  \notag \\
&=&\frac{R}{i\beta }\frac{-i}{\beta }\partial _{R}\int\limits_{0}^{2\pi
}d\varphi \exp \left[ i\beta R\cos \left( \varphi \right) \right]  \notag \\
&=&-\frac{R}{\beta ^{2}}2\pi \partial _{R}J_{0}\left( \beta R\right)  \notag
\\
&=&\pi R^{2}\left( \frac{2J_{1}\left( \beta R\right) }{\beta R}\right)
\label{68}
\end{eqnarray}%
Here the $J_{n}\left( x\right) $ are the Bessel functions of the first kind.
This is the standard result for a circular aperture\cite{FourierOptics}. .

For a slit (rectangular) aperture of width $2a_{i}$ in the $x_{i}$
direction, centered at $x_{i}=0,$ we merely have to substitute $\vec{v}%
_{1}=\left( a_{1},a_{1}\right) ,$ $\vec{v}_{2}=\left( -a_{1},a_{2}\right) ,$ 
$\vec{v}_{3}=\left( -a_{1},-a_{2}\right) ,$ $\vec{v}_{4}=\left(
a_{1},-a_{2}\right) ,$ and $\vec{v}_{5}=\vec{v}_{1}$ into (32) to obtain%
\begin{eqnarray}
A\left( \vec{x}\right) &=&-\frac{1}{\beta ^{2}}\sum_{n=1}^{N}\frac{\vec{\beta%
}\cdot \varepsilon \cdot \left( \vec{v}_{n+1}-\vec{v}_{n}\right) }{\vec{\beta%
}\cdot \left( \vec{v}_{n+1}-\vec{v}_{n}\right) }\left( \exp \left( i\vec{%
\beta}\cdot \vec{v}_{n+1}\right) -\exp \left( i\vec{\beta}\cdot \vec{v}%
_{n}\right) \right)  \notag \\
&=&\frac{1}{\beta ^{2}}\left( 
\begin{array}{c}
\frac{\beta _{2}}{\beta _{1}}\left( \exp \left( -ia_{1}\beta
_{1}+ia_{2}\beta _{2}\right) -\exp \left( i\beta _{1}a_{1}+i\beta
_{2}a_{2}\right) \right) \\ 
-\frac{\beta _{1}}{\beta _{2}}\left( \exp \left( -ia_{1}\beta
_{1}-ia_{2}\beta _{2}\right) -\exp \left( -i\beta _{1}a_{1}+i\beta
_{2}a_{2}\right) \right) \\ 
+\frac{\beta _{2}}{\beta _{1}}\left( \exp \left( ia_{1}\beta
_{1}-ia_{2}\beta _{2}\right) -\exp \left( -i\beta _{1}a_{1}-i\beta
_{2}a_{2}\right) \right) \\ 
-\frac{\beta _{1}}{\beta _{2}}\left( \exp \left( ia_{1}\beta
_{1}+ia_{2}\beta _{2}\right) -\exp \left( +i\beta _{1}a_{1}-i\beta
_{2}a_{2}\right) \right)%
\end{array}%
\right)  \notag \\
&=&\left( 2a_{1}\right) \left( 2a_{2}\right) \frac{\sin \left( \beta
_{1}a_{1}\right) }{\beta _{1}a_{1}}\frac{\sin \left( \beta _{2}a_{2}\right) 
}{\beta _{2}a_{2}}  \label{69}
\end{eqnarray}%
which\ again is the standard result\cite{FourierOptics}.

Fraunhofer diffraction patterns for arbitrary (orientable) polygons can be
calculated simply by substituting the vertex values into (32). It is
interesting to compare the patterns generated for a given set of vertices as
the vertices are reordered\ to make the polygon nonorientable.

\section{Results in Three Dimensions}

Before discussing Porods law we point out an essentially obvious but useful
fact. Since the faces of a polyhedron are themselves polygons it follows
that applying (5) to a polyhedron reduces the integral over the volume to a
sum of integrals over the areas of polygonal faces and then applying (32) to
the faces themselves reduces the volume integral to a sum of integrals over
the edges.\ These integrals can of course be evaluated exactly. The only
issue is the bookkeeping required to keep proper track of the vertices.

\subsection{Volume Element}

We use (5) to show that the volume in three dimensions of a parallelpiped
formed by three vectors $\vec{a}_{i}$ is given by $\det \left[ a_{ij}\right] 
$ where $a_{ij}$ the 3$\times $3 matrix of the components of the $\vec{a}%
_{i} $, i.e,.the first column is $a_{1i},$ the second $a_{2i}$ and the third 
$a_{3i}$ with of course, $i=1,2,3.$ If the $\vec{a}_{i}$ are infinitesimals
given by $d\vec{R}_{i}=\partial _{s_{i}}\vec{R}ds_{i}$ with no sum on $i$,
then $\det \left[ dR_{ij}\right] d^{3}s$ forms the fundamental volume
element $dv$ for integration. Expanding the right hand side of (5)\ to first
order in $\beta $ we have that the volume of a shape in three dimension is
given by%
\begin{eqnarray}
v &=&\int\limits_{V}d^{3}x  \notag \\
&=&\hat{\beta}\cdot \int\limits_{\partial V}d^{2}s\;\sqrt{g\left( \vec{s}%
\right) }\hat{n}\left( \vec{s}\right) \left( \hat{\beta}\cdot \vec{R}\left( 
\vec{s}\right) \right)  \label{70}
\end{eqnarray}%
A parallelepiped, defined by three (non coplanar) vectors $\vec{a}_{i}$, $%
i=1,2,3,$ in three dimensions has 6 parallelogram faces. The functions $\vec{%
R}_{f}\left( \vec{s}\right) $ for positions on the faces with $f$ labeling
the faces, are 
\begin{eqnarray}
\vec{R}_{1}\left( \vec{s}\right) &=&s_{1}\hat{a}_{1}+s_{2}\hat{a}_{2}\ \ 
\vec{R}_{2}\left( \vec{s}\right) =s_{1}\hat{a}_{2}+s_{2}\hat{s}_{3}\ \ \vec{R%
}_{3}\left( \vec{s}\right) =s_{1}\hat{a}_{3}+s_{2}\hat{a}_{1}  \notag \\
\vec{R}_{4}\left( \vec{s}\right) &=&\vec{a}_{3}+s_{1}\hat{a}_{1}+s_{2}\hat{a}%
_{2}\ \ \vec{R}_{5}\left( \vec{s}\right) =\vec{a}_{1}+s_{1}\hat{a}_{2}+s_{2}%
\hat{s}_{3}\ \ \vec{R}_{6}\left( \vec{s}\right) =\vec{a}_{2}+s_{1}\hat{a}%
_{3}+s_{2}\hat{a}_{1}  \label{71}
\end{eqnarray}%
The values of $s_{i}$ in each case range from 0 to the corresponding length
of the associated vector $\vec{a}_{i}.$ For example, for $f=3$, $s_{1}$
ranges from 0 to $\left\vert \vec{a}_{3}\right\vert \equiv a_{3}$ and $s_{2}$
ranges from 0 to $\left\vert \vec{a}_{1}\right\vert \equiv a_{1}.$ We assume
that the vectors $\vec{a}_{i}$ are ordered so that $\left( \vec{a}_{2}\times 
\vec{a}_{3}\right) $ points in the general direction of $\vec{a}_{1},$ i.e., 
$\vec{a}_{1}\cdot \left( \vec{a}_{2}\times \vec{a}_{3}\right) >0$ where "$%
\times $" is the standard cross product$.$ On each face $g_{f}\left( \vec{s}%
\right) $ and $\hat{n}_{f}\left( \vec{s}\right) $ are constants and so
factor out of the integrals. The outward normals $\hat{n}_{f}$ on the
opposite faces, 1 and 4, 2 and 5, 3 and 6, point in opposite directions.
Thus, the integrals over the $\vec{s}$ dependent parts of $\vec{R}_{f}\left( 
\vec{s}\right) $ completely cancel and we are left with 
\begin{eqnarray}
v &=&\left( \hat{\beta}\cdot \sqrt{g_{4}}\hat{n}_{4}\right) \left( \hat{\beta%
}\cdot \vec{a}_{3}\right) a_{1}a_{2}  \notag \\
&&+\left( \hat{\beta}\cdot \sqrt{g_{5}}\hat{n}_{5}\right) \left( \hat{\beta}%
\cdot \vec{a}_{1}\right) a_{2}a_{3}  \notag \\
&&+\left( \hat{\beta}\cdot \sqrt{g_{6}}\hat{n}_{6}\right) \left( \hat{\beta}%
\cdot \vec{a}_{2}\right) a_{3}a_{1}  \label{72}
\end{eqnarray}%
But $\hat{\beta}$ is arbitrary. Choose it to be $\hat{a}_{1}.$ Then since $%
\hat{n}_{4}=\left( \hat{a}_{1}\times \hat{a}_{2}\right) /\left\vert \hat{a}%
_{1}\times \hat{a}_{2}\right\vert $ and $\hat{n}_{6}=\left( \hat{a}%
_{3}\times \hat{a}_{1}\right) /\left\vert \hat{a}_{3}\times \hat{a}%
_{1}\right\vert $, we have $\hat{a}_{1}\cdot \hat{n}_{4}=\hat{a}_{1}\cdot 
\hat{n}_{6}=0$ and $v$ reduces to 
\begin{equation}
v=\left( \hat{a}_{1}\cdot \sqrt{g_{5}}\hat{n}_{5}\right) a_{1}a_{2}a_{3}
\label{73}
\end{equation}%
But, as shown above in the derivation of the area element in two dimensions, 
$\sqrt{g}$ is the sine of the angle between the two correponding vectors and
so equals the magnitude of the cross product of those vectors, hence $\sqrt{%
g_{5}}=\left\vert \hat{a}_{2}\times \hat{a}_{3}\right\vert $ and thus$\sqrt{%
g_{5}}\hat{n}_{5}=\hat{a}_{2}\times \hat{a}_{3}.$So finally%
\begin{eqnarray}
v &=&\hat{a}_{1}\cdot \left( \hat{a}_{2}\times \hat{a}_{3}\right)
a_{1}a_{2}a_{3}  \notag \\
&=&\vec{a}_{1}\cdot \vec{a}_{2}\times \vec{a}_{3}  \notag \\
&=&\varepsilon _{ijk}a_{1i}a_{2j}a_{3k}  \notag \\
&=&\det \left[ a_{ij}\right]  \notag \\
&=&\sqrt{\det \left[ \vec{a}_{i}\cdot \vec{a}_{j}\right] }  \label{74}
\end{eqnarray}%
Again the form in the last line is the most useful since it is coordinate
independent.

\subsection{Porods Law}

Finally, we rederive the anisotropic version of Porods law as given in the
work of Ciccariello, et. al.\cite{Ciccariello}. The anisotropic result of
course reduces to the isotropic result for spherical particles. Here we use
the notation $\vec{k}$ rather than $\vec{\beta}$ as is more common in this
context.

The intensity $I\left( \vec{k}\right) =I\left( k\hat{k}\right) =I\left( 
\frac{2\pi }{\lambda }\hat{k}\right) $ of light of wavelength $\lambda $
scattered off a particle defined by the shape function $\theta _{V}\left( 
\vec{x}\right) $ in the direction $\hat{k}=\hat{k}_{out}-\hat{k}_{in}$ where 
$\hat{k}_{in}\left( \hat{k}_{out}\right) $ is the direction of propagation
of incident(scattered) light is 
\begin{equation}
I\left( \vec{k}\right) \sim \left\vert \int d^{D}x~\exp \left[ i\vec{k}\cdot 
\vec{x}\right] \theta _{V}\left( \vec{x}\right) \right\vert ^{2}  \label{75}
\end{equation}%
To be general, for now, we start in $D$ dimensions. Porods law\cite%
{Ciccariello} follows from this in the case where the magnitude of the
scattering wavevector $k=\left\vert \vec{k}\right\vert =2\pi /\lambda $ is
large, i.e, $\lambda $ is small and so Porods law is often associated with
X-ray scattering.

To obtain Porods law, use (5) and evaluate the integral over $\partial V$ in
the limit of large $k$ using the method of stationary phase. The positions $%
\vec{\sigma}_{p}$ on the surface $\vec{R}\left( \vec{s}\right) $ where the
phase is stationary satisfy%
\begin{equation}
\left. \vec{\partial}_{s}\left( \vec{k}\cdot \vec{R}\left( \vec{s}\right)
\right) \right\vert _{\vec{s}=\vec{\sigma}_{p}}=\left. k\vec{\partial}%
_{s}\left( \hat{k}\cdot \vec{R}\left( \vec{s}\right) \right) \right\vert _{%
\vec{s}=\vec{\sigma}_{p}}=0  \label{76}
\end{equation}%
Here $\vec{\partial}_{s}=\left( \partial _{s_{1}},\partial _{s_{2}},\cdots
\right) $ is the gradient with respect to the surface coordinates and the
index $p$ runs from 1 up to the number of solutions of (76)$.$

For simplicity consider a single solution $\vec{\sigma}$ to this equation.
The argument of the exponential in the integrand can now be approximated by 
\begin{equation}
i\vec{k}\cdot \vec{R}\left( \vec{s}\right) =ik\left( \hat{k}\cdot \vec{R}%
\left( \vec{\sigma}\right) +\frac{1}{2}\left( \partial _{\sigma
_{i}}\partial _{\sigma _{j}}\hat{k}\cdot \vec{R}\left( \vec{\sigma}\right)
\right) \left( s_{i}-\sigma _{i}\right) \left( s_{j}-\sigma _{j}\right)
\right)  \label{77}
\end{equation}%
to second order in $\vec{s}-\vec{\sigma}.$ Substituting into (5) gives%
\begin{align}
& \int_{V}d^{D}x~\exp \left[ i\vec{k}\cdot \vec{x}\right]  \notag \\
& \simeq \exp \left[ ik\hat{k}\cdot \vec{R}\left( \vec{\sigma}\right) \right]
\frac{\hat{k}}{ik}\cdot \int d^{D-1}s\ \sqrt{g\left( \vec{s}\right) }\hat{n}%
\left( \vec{s}\right) \exp \left[ \frac{ik}{2}\left( \partial _{\sigma
_{i}}\partial _{\sigma _{j}}\hat{k}\cdot \vec{R}\left( \vec{\sigma}\right)
\right) \left( s_{i}-\sigma _{i}\right) \left( s_{j}-\sigma _{j}\right) %
\right]  \notag \\
& \simeq \exp \left[ ik\hat{k}\cdot \vec{R}\left( \vec{\sigma}\right) \right]
\left( \frac{\hat{k}}{ik}\cdot \hat{n}\left( \vec{\sigma}\right) \right) 
\frac{\sqrt{g\left( \vec{\sigma}\right) }\pi ^{\left( D-1\right) /2}}{\sqrt{%
\det \left[ -\frac{ik}{2}\partial _{\sigma _{i}}\partial _{\sigma _{j}}\hat{k%
}\cdot \vec{R}\left( \vec{\sigma}\right) \right] }}  \label{78}
\end{align}%
where the determinant in the denominator is taken over the $i,j$ indices.
Using the fact that $\det \left[ -\frac{ik}{2}\partial _{\sigma
_{i}}\partial _{\sigma _{j}}\hat{k}\cdot \vec{R}\left( \vec{\sigma}\right) %
\right] =k^{D-1}\det \left[ -\frac{i}{2}\partial _{\sigma _{i}}\partial
_{\sigma _{j}}\hat{k}\cdot \vec{R}\left( \vec{\sigma}\right) \right] $ we get%
\begin{equation}
\int_{V}d^{D}x~\exp \left[ i\vec{k}\cdot \vec{x}\right] \sim \frac{1}{k}%
\frac{1}{k^{\left( D-1\right) /2}}=\frac{1}{k^{\left( D+1\right) /2}}
\label{79}
\end{equation}%
and so 
\begin{equation}
I\left( \vec{k}\right) \sim \frac{1}{k^{D+1}}  \label{80}
\end{equation}%
Thus in 3 dimensions $I\left( \vec{k}\right) $ scales as $1/k^{4}$ for large 
$k.$ This is the standard isotropic statement of Porods law\cite{Porod}. But
as pointed out by Ciccariello, $\det \left[ \partial _{\sigma _{i}}\partial
_{\sigma _{j}}\hat{k}\cdot \vec{R}\left( \vec{\sigma}\right) \right] $ is
proportional to the Gaussian curvature (in 3 dimensions) of the surface $%
\partial V.$ In general for an arbitrary shaped particle there will be
multiple stationary phase points, i.e., $\vec{\sigma}_{1},\vec{\sigma}%
_{2},\cdots $ which must be summed to get the complete amplitude. There are
also issues with positive and negative curvature which must be carefully
considered as well along with how the handle the case when the curvature
vanishes\cite{Ciccariello}.

\section{Conclusion}

We have shown how a simple idea, that of combining Gauss's law with the
Fourier transform provides alternative solutions and/or derivations of many
different classical results in physics and mathematics. No doubt there are
many other problems and proofs to which this idea can be applied.


\begin{thebibliography}{99}
\bibitem{SCO} Simply connected here means that the shape is a single
connected area in two dimensions, a single connected volume in three
dimensions and not multiple disconnected shapes. By orientable here we mean
simply that the boundary manifold of the shape (the curve enclosing an area
in two dimensions, the surface enclosing a volume in three dimensions, etc.)
does not intersect itself.
\end{thebibliography}
\end{document}